\documentclass[nofonts]{fairmeta}
\usepackage{hyperref}       % hyperlinks
\usepackage{url}            % simple URL typesetting
\usepackage{booktabs}       % professional-quality tables
\usepackage{amsfonts}       % blackboard math symbols
\usepackage{nicefrac}       % compact symbols for 1/2, etc.
\usepackage{microtype}      % microtypography
\usepackage{xcolor}         % colors
\usepackage[english]{babel}
\usepackage{amsmath}
\usepackage{amssymb}
\usepackage{booktabs}
\usepackage{algorithm}
\usepackage{algorithmic}

\usepackage{graphicx}
\usepackage{multirow}
\usepackage{tikz}
\usetikzlibrary{positioning, fit, backgrounds, arrows.meta, shapes.geometric, calc}

\title{Synthetic Data from Cross-Domain Events for Large-Scale Recommendation Systems}

\author[1,*]{Xiangyu Wang}
\author[1,*]{Yawen He}
\author[1,*]{Shivendra Pratap Singh}
\author[1,*]{Han Huang}
\author[1,*]{Mengtong Hu}
\author[1,*]{Sharath Ciddu}
\author[1,*]{Yi-Hsuan Hsieh}
\author[1,*]{Erik Groving}
\author[1,*]{Yi Ding}
\author[1,*]{Jieming Di}
\author[1,*]{Tony Wang}
\author[1,*]{Min Yun}
\author[1,*]{Xiaoyu Chen}
\author[1,*]{Ling Leng}
\author[1,*]{Rob Malkin}

\affiliation[1]{Meta}
\contribution[*]{Work done at Meta}

\abstract{Large-scale recommendation systems operate across diverse domains, yet face the challenges of data sparsity and noisy implicit feedback. Traditional approaches mitigate this by model-specific knowledge distillation from source domains to a target domain. Inspired by the transformative success of synthetic data generation in large language model (LLM), we introduce {\underline{S}ynthetic \underline{C}ross-domain \underline{A}ugmentation and \underline{L}earning for \underline{R}ecommendation (SCALR)}, a framework that generates synthetic user-item interaction events for a target recommendation domain by leveraging observed events from a source domain. SCALR decomposes cross-domain learning into two modular stages. First it translates observed user event in source domains by formulating event generation as estimating the likelihood that a user would interact with a target-domain item, conditioned on their observed interactions in a source domain. Second, downstream models train on these synthetic events as cross-domain learning objectives where synthetic events augment the target domain's training data in a model-agnostic manner. Our approach yields statistically significant improvements in online A/B tests on an industrial recommendation platform. To the best of our knowledge, this is among the first works to explicitly frame cross-domain event transfer as synthetic data generation for recommendation systems.}

\correspondence{\email{<wangxy,yawenhe,shiven>@meta.com}}

\begin{document}

%%
%% The abstract is a short summary of the work to be presented in the
%% article.
% \begin{abstract}
% Large-scale recommendation systems operate across diverse domains, yet face the challenges of data sparsity and noisy implicit feedback. Traditional approaches mitigate this by model-specific knowledge distillation from source domains to a target domains. Inspired by the transformative success of synthetic data generation in large language model (LLM), we introduce {\underline{S}ynthetic \underline{C}ross-domain \underline{A}ugmentation and \underline{L}earning for \underline{R}ecommendation (SCALR)}, a framework that generates synthetic user-item interaction events for a target recommendation domain by leveraging observed events from a source domain. SCALR decomposes cross-domain learning into two modular stages. First it translates observed user event in source domains by formulating event generation as estimating the likelihood that a user would interact with a target-domain item, conditioned on their observed interactions in a source domain. Second, downstream models train on these synthetic events as a cross-domain learning objective where synthetic events augment the target domain's training data in a model-agnostic manner. Our approach yields statistically significant improvements in online A/B tests on an industrial recommendation platform. To the best of our knowledge, this is among the first works to explicitly frame cross-domain event transfer as synthetic data generation for recommendation systems.
% \end{abstract}

\maketitle

% %%
% %% Keywords. The author(s) should pick words that accurately describe
% %% the work being presented. Separate the keywords with commas.
%\keywords{synthetic data, recommendation systems, cross-domain learning, data augmentation, conversion prediction, event generation}
%% A "teaser" image appears between the author and affiliation
%% information and the body of the document, and typically spans the
%% page.
% \begin{teaserfigure}
%   \includegraphics[width=0.5\textwidth]{SynRec_illustration.png}
%   \caption{Illustration of the recommendation system with source domain (post / website) and target domain (product).}
%   \Description{}
%   \label{fig:teaser}
% \end{teaserfigure}

% \received{20 February 2007}
% \received[revised]{12 March 2009}
% \received[accepted]{5 June 2009}

%%
%% This command processes the author and affiliation and title
%% information and builds the first part of the formatted document.
% \maketitle

% ==============================================================================
\section{Introduction}
\label{sec:introduction}
% ==============================================================================

The use of synthetic data to train machine learning models has become one of the most impactful ideas in modern AI. In the domain of large language models (LLMs), synthetic data has been widely adopted across pre-training, instruction tuning, alignment, and knowledge distillation, consistently demonstrating that high-quality synthetic examples can match or even surpass the organically collected data~\cite{gunasekar2023textbooks,alpaca2023,liu2024bestpractices}.

Despite this success, the application of synthetic data to recommendation systems has received comparatively limited attention as an explicit paradigm. Existing uses of synthetic data in this space have focused on either privacy preservation~\cite{slokom2018comparing,drechsler2011empirical} or adversarial training~\cite{wang2017irgan}, rather than on generating cross-domain training events to address data sparsity---the focus of our work.

Recommendation systems, which power content feeds and product suggestions at scale, face a fundamental challenge: the training data consists of implicit feedback (clicks, conversions, engagements) that is inherently noisy, sparse, and heavily skewed. Existing approaches to address this sparsity fall into two categories. Cross-domain recommendation methods~\cite{zhu2021crossdomain,hu2018conet,man2017cross} transfer knowledge from a data-rich source domain to the target domain, but they do so at the \emph{model level}---sharing user embeddings, latent factors, or network parameters through joint training or learned mapping functions. These methods have several practical limitations. First, \emph{architecture coupling}: because the source domain signal is absorbed into shared model parameters, the transfer mechanism is entangled with the model architecture. Adopting a different downstream model (e.g., switching from a two-tower model to a transformer-based ranker) requires re-designing the cross-domain components, which limits adoption in production systems where model architectures evolve frequently. Second, \emph{training coupling}: source and target domain models must typically be co-trained or sequentially fine-tuned, meaning that changes to the source domain data or model require retraining the entire pipeline end-to-end. Third, \emph{scalability to multiple consumers}: when multiple downstream models (e.g., ranking, calibration, bidding) could benefit from the same source domain signal, each model must independently implement its own cross-domain transfer mechanism, duplicating effort and introducing inconsistencies. Data augmentation techniques~\cite{wang2021counterfactual,yao2021selfsupervised,xie2022contrastive} take a different approach but operate \emph{within a single domain}, creating perturbed views of existing interactions through item masking, sequence cropping, or counterfactual substitution. These methods improve representation learning but do not bring in behavioral signal from outside the target domain.

\begin{figure*}
\centering
  \includegraphics[width=0.7\linewidth]{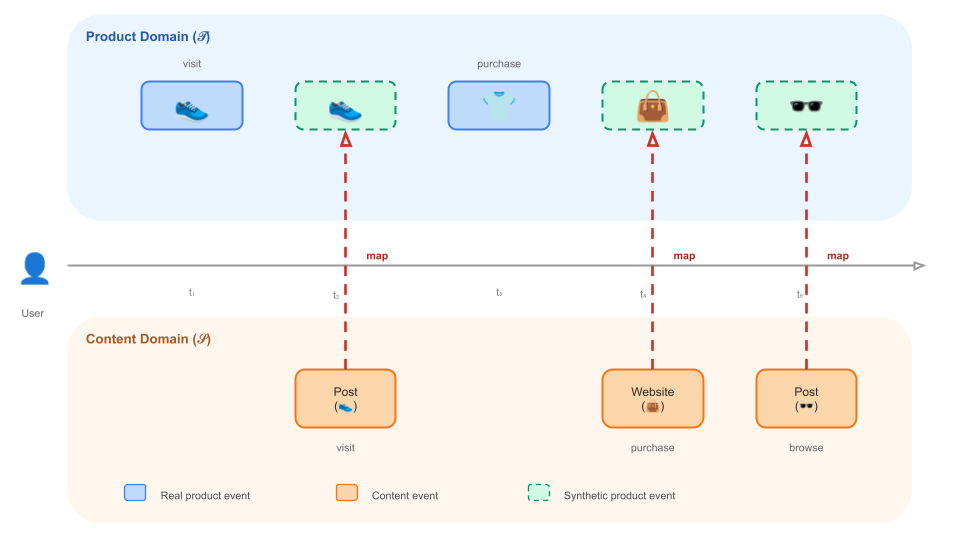}
  \caption{Illustration of the recommendation system with source domain (post / website) and target domain (product).}
  \label{fig:teaser}
\end{figure*}

Our work takes a fundamentally different approach that decomposes cross-domain learning into two distinct stages. In the first stage, \emph{synthetic data generation}, source domain events are translated into concrete target domain training examples before any model training begins. In the second stage, \emph{cross-domain learning}, the downstream model is trained on the combination of original and synthetic events through a weighted objective that controls the influence of each data source. While this second stage is itself a form of cross-domain learning, it differs from prior cross-domain recommendation methods: the cross-domain alignment has already been resolved during data generation, so the learning stage operates entirely within the target domain's data format. The downstream model never needs to know that some training events originated from another domain as it consumes synthetic events in the same format as original events. This separation makes the approach \emph{model-agnostic}: any recommendation architecture benefits without architectural modification. It is also \emph{modular}: the synthetic data generation pipeline can be iterated on independently of the recommendation model, and multiple downstream models can reuse the same synthetic dataset. Furthermore, because synthetic events from different source domains are all translated into the same target domain format, the framework naturally supports \emph{multi-source composition}: synthetic data from multiple source domains (e.g., browsing behavior, content engagement, search queries) can be combined simply by concatenating the generated events, without any additional architectural or algorithmic complexity. Finally, because the synthetic events are generated through a probabilistic translation rather than copied directly from source domain logs, they also offer a degree of \emph{privacy preservation}: the generated events do not correspond to any single user's raw activity record, which may reduce privacy risks compared to directly sharing or transferring raw cross-domain interaction data~\cite{jordon2022synthetic}. By contrast, traditional cross-domain methods~\cite{zhu2021crossdomain,hu2018conet} must embed the cross-domain transfer mechanism directly into the model, coupling data transfer and model training into a single, architecture-specific procedure.

In this paper, we introduce \textbf{SCALR}, a framework that casts cross-domain event transfer as synthetic data learning for recommendation systems. User interaction events in a source domain $\mathcal{S}$ (e.g., browsing behavior, content engagement) serve as raw material from which synthetic training events are generated for a target domain $\mathcal{T}$. We formulate event generation as a conditional probability estimation problem, present a frequency-based approach leveraging co-occurrence statistics, and further explore a model-based method using learned embeddings. We show that probabilistic sampling from the translation distribution outperforms deterministic top-$K$ selection. In online A/B tests, SCALR yields statistically significant improvements in core business metrics. To our knowledge, this is among the first works to explicitly frame cross-domain event transfer as synthetic data generation for recommendation systems. 

The remainder of this paper is organized as follows. Section~\ref{sec:related} surveys related work on synthetic data in LLMs, synthetic data in recommendation systems, cross-domain recommendation, and data augmentation for recommendation systems. Section~\ref{sec:method} presents the SCALR framework in detail: we formalize the problem setting with notation grounded in the recommendation domain, formulate the synthetic data generation objective as a conditional probability estimation problem, describe a frequency-based estimation approach, and formulate the downstream training as a cross-domain learning objective. Section~\ref{sec:comparison} provides a systematic comparison between synthetic data paradigms in LLMs and recommendation systems. Section~\ref{sec:experiments} reports experimental results on an industrial platform, including online A/B tests and an analysis of generation strategies. Finally, Section~\ref{sec:discussion} discusses implications and limitations, and Section~\ref{sec:conclusion} concludes.

% ==============================================================================
\section{Related Work}
\label{sec:related}
% ==============================================================================

\paragraph{Synthetic data generation and usage in LLMs.}
Synthetic data has become a central paradigm in large language model training. Gunasekar et al.~\cite{gunasekar2023textbooks} demonstrated with Phi-1 that a 1.3B-parameter model trained on synthetically generated ``textbook-quality'' data can match GPT-3.5-level coding performance, establishing that data quality can substitute for model scale. In the alignment setting, the Alpaca project~\cite{alpaca2023} showed that instruction-following capabilities can be distilled via synthetic instruction-response pairs, while Orca~\cite{mukherjee2023orca} extended this with detailed reasoning traces. Liu et al.~\cite{liu2024bestpractices} distilled key lessons from this body of work, finding that distribution alignment between synthetic and real data, quality filtering, and careful control of the synthetic-to-real data ratio are critical for success. While these principles translate conceptually to recommendation systems, the underlying data modality is fundamentally different: LLM synthetic data operates over continuous token sequences with a shared vocabulary, whereas recommendation systems require generating \emph{discrete user-item interaction events} across disjoint item catalogs, subject to heavy-tailed distributions and implicit feedback noise.

\paragraph{Synthetic data in recommendation systems.}
While synthetic data has transformed LLM training, its use in recommendation systems has been limited and motivated by different goals. Slokom~\cite{slokom2018comparing} applied CART-based synthetic data generation~\cite{drechsler2011empirical} to collaborative filtering, producing partially or fully synthetic user profiles to replace privacy-sensitive values---enabling recommendation system evaluation without exposing real user data. This line of work treats synthetic data as a \emph{privacy mechanism} rather than a means to improve model performance. In a different vein, Wang et al.~\cite{wang2017irgan} proposed IRGAN, a minimax adversarial framework for information retrieval in which a generative model learns to select pseudo-relevant items from a candidate pool to fool a discriminative retrieval model. Crucially, IRGAN does not \emph{generate} new data---its generator \emph{selects existing items} as pseudo-relevant samples, operating entirely within a single domain. It neither creates novel interaction events nor transfers behavioral signals across domains. Our work differs from both directions: we generate synthetic training events by translating cross-domain user interactions into the target domain's format, with the explicit goal of augmenting training data to improve downstream recommendation performance.

\paragraph{Cross-domain recommendation.}
Cross-domain recommendation (CDR) addresses data sparsity by transferring knowledge from a data-rich source domain to a sparse target domain~\cite{zhu2021crossdomain}. Representative approaches include Collective Matrix Factorization~\cite{singh2008relational}, which shares user latent factors across domain-specific interaction matrices; EMCDR~\cite{man2017cross}, which learns an explicit mapping between source and target user embeddings; and CoNet~\cite{hu2018conet}, which enables bidirectional transfer through cross-stitch connections in dual deep networks. More recent work such as DisenCDR~\cite{cao2022disencdr} disentangles domain-shared from domain-specific representations to enable more targeted transfer. However, CDR methods universally operate at the \emph{representation level}---they share parameters, embeddings, or latent factors within a jointly trained model. No prior work in CDR has framed the transfer as \emph{synthetic data generation}, where source domain interactions are translated into concrete training events for the target domain, decoupled from any specific downstream model.

\paragraph{Data augmentation for recommendation systems.}
Within-domain augmentation strategies for recommendation systems include counterfactual sequence augmentation~\cite{wang2021counterfactual}, self-supervised contrastive methods that create augmented views through item masking or feature dropout~\cite{yao2021selfsupervised,xie2022contrastive}, and generative approaches that hallucinate pseudo-prior interactions~\cite{lee2022augmenting}. More recently, LLMs have been used to augment recommendation signals: LLMRec~\cite{wei2024llmrec} generates synthetic user-item interactions for graph-based recommendation, and TallRec~\cite{bao2023tallrec} adapts LLMs for recommendation through instruction tuning. While these methods enrich training data, they operate within a single domain and do not leverage cross-domain behavioral signals. Our work is distinct in that it generates synthetic events by translating interactions from an entirely different domain---casting cross-domain event transfer as a synthetic data generation problem, analogous to how teacher model outputs serve as synthetic training data in the LLM setting.

% ==============================================================================
\section{The SCALR Framework}
\label{sec:method}
% ==============================================================================

We now present SCALR, our framework for synthetic data learning in recommendation systems. We begin by formalizing the problem with concrete notation grounded in the recommendation setting, then describe our synthetic data generation procedure, and finally formulate the downstream training as cross-domain learning.

\begin{figure*}[t]
  \centering
  \resizebox{0.85\textwidth}{!}{\begin{tikzpicture}[
    >=Stealth,
    font=\sffamily,
    % Core Color Definitions
    darkblue/.style={color=blue!30!black},
    topborder/.style={color=purple!40},
    topfill/.style={fill=purple!4},
    botborder/.style={color=green!50!black},
    botfill/.style={fill=green!4},
    panelborder/.style={draw=gray!50, fill=white, rounded corners=12pt, thick},
    % Domain and Item Box Styles
    sourcebg/.style={fill=orange!10, draw=orange!30, rounded corners=8pt, thick},
    targetbg/.style={fill=blue!5, draw=blue!30, rounded corners=8pt, thick},
    observeditem/.style={draw=orange!60, fill=white, rounded corners=4pt, inner sep=6pt, font=\small, thick, align=center},
    syntheticitem/.style={draw=green!60!black, fill=green!10, dashed, rounded corners=4pt, inner sep=6pt, font=\small, thick, align=center},
    % Feature Boxes
    featuser/.style={draw=blue!40, fill=blue!10, rounded corners=4pt, minimum height=1.2cm, minimum width=1.1cm, align=center, font=\scriptsize, text=blue!70!black},
    featitem/.style={draw=orange!40, fill=orange!10, rounded corners=4pt, minimum height=1.2cm, minimum width=1.1cm, align=center, font=\scriptsize, text=orange!70!black},
    featctx/.style={draw=gray!40, fill=gray!10, rounded corners=4pt, minimum height=1.2cm, minimum width=0.9cm, align=center, font=\scriptsize, text=gray!70!black},
    featy/.style={draw=green!60!black, fill=green!10, rounded corners=4pt, minimum height=1.2cm, minimum width=0.8cm, align=center, font=\scriptsize\bfseries, text=black, thick},
    % Cross-Domain Learning Boxes
    origdata/.style={draw=blue!50, fill=blue!5, rounded corners=6pt, minimum height=1.4cm, minimum width=4.8cm, align=center, text=blue!80!black, thick},
    syndata/.style={draw=green!60!black, fill=green!10, dashed, rounded corners=6pt, minimum height=1.4cm, minimum width=4.8cm, align=center, text=green!50!black, ultra thick},
    model/.style={draw=none, fill=blue!15!black, rounded corners=8pt, text=white, align=center, minimum height=2.2cm, minimum width=3.8cm, font=\bfseries},
    pred/.style={draw=blue!50, fill=blue!5, rounded corners=6pt, minimum height=1.4cm, minimum width=2.8cm, align=center, text=blue!80!black, thick, font=\bfseries}
]

% ==========================================
% TITLE
% ==========================================
\node[font=\Large\bfseries, color=blue!20!black, align=center] at (6.65, 3.2) {SCALR: Proposed Flow};

% ==========================================
% TOP SECTION: ① Synthetic Data Generation
% ==========================================
\begin{scope}[on background layer]
    \node[topfill, draw=purple!30, thick, rounded corners=15pt, minimum width=18.6cm, minimum height=4.8cm] (top_bg) at (6.65, 0) {};
\end{scope}
\node[anchor=north west, font=\bfseries\color{purple!70!black}, yshift=-0.2cm, xshift=0.3cm] at (top_bg.north west) {\tikz[baseline=-0.7ex]{\node[circle,draw,inner sep=0.5pt,font=\scriptsize]{1};} Synthetic Data Generation};

% Event Generation Panel
\node[panelborder, minimum width=8.7cm, minimum height=3.6cm] (event_panel) at (2.1, -0.2) {};
\node[anchor=north west, font=\small\bfseries, text=black!80, yshift=-0.2cm, xshift=0.2cm] at (event_panel.north west) {Event Generation};

% Source Domain Setup
\node[sourcebg, minimum width=3.5cm, minimum height=2.4cm] (source_box) at ([yshift=-0.2cm, xshift=-2.2cm]event_panel.center) {};
\node[anchor=north west, font=\scriptsize\bfseries, text=orange!80!black, yshift=-0.1cm, xshift=0.1cm] at (source_box.north west) {Source Domain ($\mathcal{S}$)};
\node[anchor=south, font=\scriptsize\itshape, text=orange!80!black, yshift=0.1cm] at (source_box.south) {observed event};
\node[observeditem, minimum width=2.4cm, minimum height=0.9cm] (post) at (source_box.center) {\textbf{Post} ({\color{blue!60!cyan} shoes})};

% Target Domain Setup
\node[targetbg, minimum width=3.5cm, minimum height=2.4cm] (target_box) at ([yshift=-0.2cm, xshift=2.2cm]event_panel.center) {};
\node[anchor=north west, font=\scriptsize\bfseries, text=blue!80!black, yshift=-0.1cm, xshift=0.1cm] at (target_box.north west) {Target Domain ($\mathcal{T}$)};
\node[anchor=south, font=\scriptsize\itshape, text=blue!80!black, yshift=0.1cm] at (target_box.south) {synthetic event};
\node[syntheticitem, minimum width=2.4cm, minimum height=0.9cm] (syn_item) at (target_box.center) {\textbf{Product (shoes)}};

% Map Arrow
\draw[dashed, red!70!black, thick, -{Triangle[open, color=red!70!black, length=3mm, width=2.5mm]}] (post) -- node[above, font=\scriptsize\bfseries, yshift=-0.1cm] {map} node[below, font=\scriptsize\bfseries, yshift=0.1cm] {$\hat{P}(i|j)$} (syn_item);

% Feature & Row Generation Panel
\node[panelborder, minimum width=8.7cm, minimum height=3.6cm] (feat_panel) at (11.2, -0.2) {};
\node[anchor=north west, font=\small\bfseries, text=black!80, yshift=-0.2cm, xshift=0.2cm] at (feat_panel.north west) {Feature \& Row Generation};

% Elements inside Feature Panel (Shifted slightly left to make room for arrow)
\node[syntheticitem, minimum width=2cm, minimum height=1.3cm] (syn_event2) at ([yshift=-0.2cm, xshift=-2.9cm]feat_panel.center) {Synthetic\\Event \scriptsize $(u, i)$};

% Feature blocks (Shifted slightly right to make room for arrow)
\begin{scope}[shift={([yshift=-0.2cm, xshift=0.1cm]feat_panel.center)}]
    \node[featuser] (f_user) at (0,0) {user\\feat};
    \node[featitem] (f_item) at (1.3,0) {item\\feat};
    \node[featctx] (f_ctx) at (2.5,0) {context\\feat};
    \node[featy] (f_y) at (3.5,0) {$\tilde{y}=0/1$};
\end{scope}

% Feature Extraction Arrow (Now longer due to the shifts above)
\draw[->, gray!80, thick, -Stealth, line width=1pt] (syn_event2) -- node[above, font=\scriptsize\itshape, text=gray!80!black, align=center, yshift=-0.1cm] {Feature\\Extraction} (f_user);

% Shifted this label slightly right to stay centered under the feature blocks
\node[font=\scriptsize\itshape, text=gray!60!black] at ([yshift=-1.4cm, xshift=1.8cm]feat_panel.center) {Training Data Row $(x, \tilde{y}å)$};

% Connecting arrow between main panels
\draw[->, gray!70, thick, line width=1.5pt, -{Stealth[length=3mm, width=2.5mm]}] (event_panel.east) -- (feat_panel.west);

% ==========================================
% BOTTOM SECTION: ② Cross-Domain Learning
% ==========================================
\begin{scope}[on background layer]
    \node[botfill, draw=green!40, thick, rounded corners=15pt, minimum width=18.6cm, minimum height=4.5cm] (bot_bg) at (6.65, -5.5) {};
\end{scope}
\node[anchor=north west, font=\bfseries\color{green!50!black}, yshift=-0.2cm, xshift=0.3cm] at (bot_bg.north west) {\tikz[baseline=-0.7ex]{\node[circle,draw,inner sep=0.5pt,font=\scriptsize]{2};} Cross-Domain Learning};

% Datasets
\node[origdata] (d_t) at ([yshift=0.8cm, xshift=-5cm]bot_bg.center) {Original Target Data D\_T \\ \footnotesize $(u, i, y)$ events};
\node[syndata] (d_syn) at ([yshift=-0.8cm, xshift=-5cm]bot_bg.center) {Synthetic Data D\_syn \\ \footnotesize $(u, i, \tilde{y})$ from SCALR};

% Model & Output
\node[model] (model) at ([yshift=0cm, xshift=0.5cm]bot_bg.center) {Model Training \\ $f_\theta$};
\node[pred] (predictions) at ([yshift=0cm, xshift=5.5cm]bot_bg.center) {Predictions \\ $\hat{y}$};

% Arrows
\draw[->, blue!60!black, thick, -{Stealth[length=3mm, width=2.5mm]}, line width=1.2pt] (d_t.east) -- (d_t.east -| model.west);
\draw[->, green!60!black, thick, -{Stealth[length=3mm, width=2.5mm]}, line width=1.2pt] (d_syn.east) -- (d_syn.east -| model.west);
\draw[->, darkblue, thick, -{Stealth[length=3mm, width=2.5mm]}, line width=1.2pt] (model.east) -- (predictions.west);

% Loss Equation
\node[below=0.2cm of model, font=\footnotesize, text=black!80] {$\mathcal{L} = \mathcal{L}_{\text{orig}} + \lambda \cdot \mathcal{L}_{\text{syn}}$};

\end{tikzpicture}
}
\caption{Illustration of Proposed Flow}
 \end{figure*}
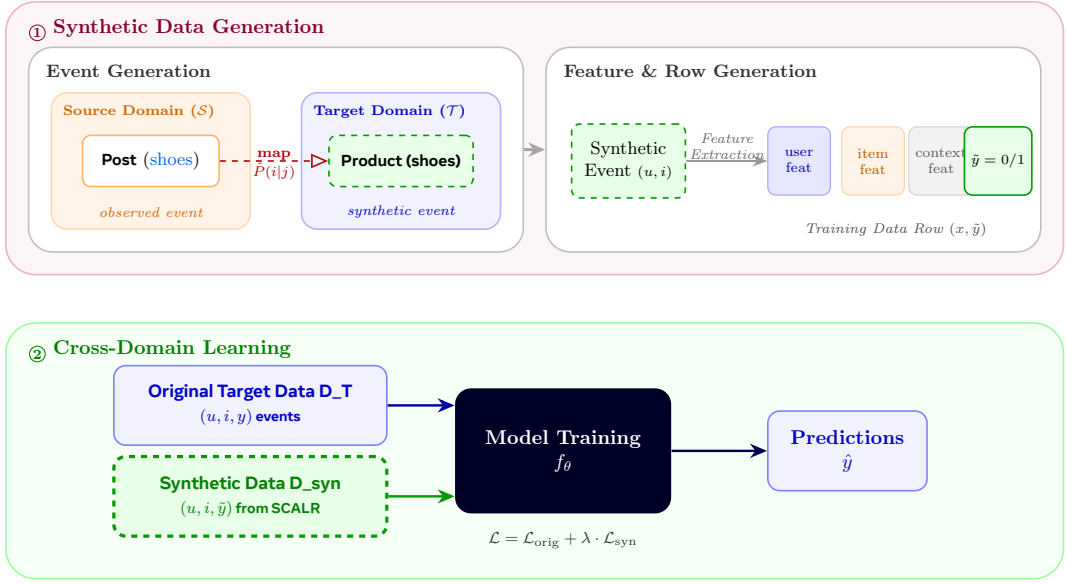

\subsection{Problem Setting and Notation}
\label{sec:notation}

Consider a platform where the goal is to predict whether a user $u$ will convert after being shown an item $i$. The training data for this \emph{target domain} $\mathcal{T}$ consists of observed $\langle u, i \rangle$ interaction events---specifically, user-item pairs where a conversion occurred. In practice, conversion events are extremely sparse: conversion rates are typically below 1\%, and many items observe only tens to hundreds of conversions per day, making it difficult to train accurate conversion prediction models.

Meanwhile, the same platform observes abundant user behavior in other \emph{source domains} $\mathcal{S}$. For example, a social platform records $\langle u, j \rangle$ interactions representing users visiting websites, engaging with content, or browsing product pages. These source domain events are orders of magnitude more plentiful than conversion events, and they carry rich signal about user preferences and intent.

Formally, let $\mathcal{U}$ denote the shared user space across domains. Let $\mathcal{I}_{\mathcal{S}}$ and $\mathcal{I}_{\mathcal{T}}$ denote the item spaces for the source and target domains, respectively. In the recommendation setting, $\mathcal{I}_{\mathcal{T}}$ is the set of recommendations and $\mathcal{I}_{\mathcal{S}}$ is the set of websites, content pages, or products in the source surface. In general, $\mathcal{I}_{\mathcal{S}} \cap \mathcal{I}_{\mathcal{T}} = \emptyset$, such that the two domains have disjoint item catalogs. We denote the observed interaction logs as:
\begin{align}
\mathcal{D}_{\mathcal{S}} &= \{(u_k, j_k) : u_k \in \mathcal{U},\; j_k \in \mathcal{I}_{\mathcal{S}}\}_{k=1}^{N_{\mathcal{S}}} \\
\mathcal{D}_{\mathcal{T}} &= \{(u_k, i_k) : u_k \in \mathcal{U},\; i_k \in \mathcal{I}_{\mathcal{T}}\}_{k=1}^{N_{\mathcal{T}}}
\end{align}
where $N_{\mathcal{S}} \gg N_{\mathcal{T}}$. The central question is: \emph{can we leverage the abundant source domain events $\mathcal{D}_{\mathcal{S}}$ to synthesize additional training events for the target domain $\mathcal{T}$?}

\subsection{Synthetic Data Generation}
\label{sec:generation}

The goal of synthetic data generation is to produce target domain events from source domain observations. We denote this as: given a source domain event $\langle u, j_{\mathcal{S}} \rangle$, generate a target domain event $\langle u, i_{\mathcal{T}} \rangle$. Since the user identity is shared across domains, the core quantity to estimate is the \emph{item translation distribution}:
\begin{equation}
P(i_{\mathcal{T}} \mid u, j_{\mathcal{S}})
\label{eq:item_translation}
\end{equation}
which defines a probability distribution over all target domain items $i \in \mathcal{I}_{\mathcal{T}}$, conditioned on user $u$ and the source item $j$ that the user interacted with. In the recommendation example, this captures: given that user $u$ visited a merchant's page $j$, how likely is $u$ to interact with each product $i$ offered by that merchant?

The formulation in Equation~\ref{eq:item_translation} defines a \emph{distribution}, not a single prediction. Rather than seeking a single best target item $i^* = \arg\max_i P(i_{\mathcal{T}} \mid u, j_{\mathcal{S}})$, we maintain the full distribution over the target item space, as the choice between deterministic selection and stochastic sampling has significant implications for synthetic data quality (Section~\ref{sec:topk_vs_sampling}). Importantly, the translation distribution is not learned by a generative model but estimated from observed co-occurrence statistics in the target domain via a frequency-based approach. We later extend this to a model-based estimator using learned embeddings (Section~\ref{sec:discussion}).

\paragraph{Frequency-based translation}
We propose a frequency-based approach as an efficient and scalable method for estimating Equation~\ref{eq:item_translation}. The key idea is to leverage co-occurrence statistics from the set of \emph{overlapping users} $\mathcal{U}_{\text{overlap}} = \{u : \exists\, (u, j, \cdot) \in \mathcal{D}_{\mathcal{S}} \;\text{and}\; \exists\, (u, i, \cdot) \in \mathcal{D}_{\mathcal{T}}\}$ who are active in both domains. For these users, we compute the empirical item translation probability:
\begin{equation}
\hat{P}(i_{\mathcal{T}} \mid j_{\mathcal{S}}) = \frac{\sum_{u \in \mathcal{U}_{\text{overlap}}} \mathbf{1}[(u, j) \in \mathcal{D}_{\mathcal{S}}] \cdot \mathbf{1}[(u, i) \in \mathcal{D}_{\mathcal{T}}]}{\sum_{u \in \mathcal{U}_{\text{overlap}}} \mathbf{1}[(u, j) \in \mathcal{D}_{\mathcal{S}}]}
\label{eq:freq}
\end{equation}
where $\mathbf{1}[\cdot]$ is the indicator function.

\paragraph{Generation algorithm.}
Given the estimated translation distribution, Algorithm~\ref{alg:scalr} describes the generation pipeline. For each source domain event, we compute the translation distribution over candidate target items, draw $K$ target items from this distribution, and produce synthetic events. In general, each synthetic event can be assigned a confidence weight $w_k = g(\hat{P}(i_k \mid u, j))$ for some weighting function $g$, allowing the downstream model to differentiate between high- and low-confidence synthetic examples. In our current implementation, we use a uniform weight $w_k = 1$ for all sampled events, since the sampling distribution itself already encodes the translation probability---higher-probability items are drawn more frequently, so the aggregate synthetic dataset naturally reflects the estimated distribution without per-event reweighting.

\begin{algorithm}[ht]
\caption{SCALR: Synthetic Event Generation}
\label{alg:scalr}
\begin{algorithmic}[1]
\REQUIRE Source domain events $\mathcal{D}_{\mathcal{S}}$, translation distribution $\hat{P}(i_{\mathcal{T}} \mid u, j_{\mathcal{S}})$, generation budget $K$ per source event, weighting function $g$
\ENSURE Synthetic target domain events $\mathcal{D}_{\text{syn}}$
\STATE $\mathcal{D}_{\text{syn}} \leftarrow \emptyset$
\FOR{each $(u, j) \in \mathcal{D}_{\mathcal{S}}$}
    \STATE Compute $\hat{P}(i_{\mathcal{T}} \mid u, j_{\mathcal{S}})$ for candidate items $i \in \mathcal{I}_{\mathcal{T}}$
    \FOR{$k = 1, \ldots, K$}
        \STATE Draw $i_k \sim \hat{P}(\cdot \mid u, j_{\mathcal{S}})$
        \STATE $\mathcal{D}_{\text{syn}} \leftarrow \mathcal{D}_{\text{syn}} \cup \{(u, i_k, w_k)\}$ where $w_k = g(\hat{P}(i_k \mid u, j))$
    \ENDFOR
\ENDFOR
\RETURN $\mathcal{D}_{\text{syn}}$
\end{algorithmic}
\end{algorithm}

\paragraph{Top-k sampling versus deterministic selection.}
A natural alternative to stochastic sampling in Algorithm~\ref{alg:scalr} is \emph{deterministic top-$K$ selection}: for each source event, select the $K$ target items with the highest $\hat{P}(i_{\mathcal{T}} \mid u, j_{\mathcal{S}})$. While top-$K$ selection maximizes the per-event translation probability, it produces a synthetic dataset with low item diversity, given that the same high-probability items are selected repeatedly across source events, concentrating the synthetic data on a narrow slice of the target item catalog.

This trade-off is similar to the decoding strategy choice in language generation. Holtzman et al.~\cite{holtzman2020curious} showed that greedy and top-$K$ decoding lead to degenerate, repetitive text, while \emph{nucleus sampling} (top-$p$) produces more natural and diverse outputs by sampling from the high-probability portion of the distribution. Fan et al.~\cite{fan2018hierarchical} similarly demonstrated that stochastic sampling with appropriate truncation yields more diverse and coherent story generation than beam search. The underlying principle is that maximizing likelihood at each step does not maximize the quality of the \emph{aggregate} output. Instead, diversity and distributional coverage matter. In the synthetic data context specifically, Gudibande et al.~\cite{gudibande2023false} and Liu et al.~\cite{liu2024bestpractices} found that diversity in synthetic training data is a key driver of downstream performance, and that over-reliance on high-confidence examples leads to narrow coverage and degraded generalization.

The same principle applies to synthetic event generation for recommendation systems. Stochastic sampling from $\hat{P}(\cdot \mid u, j_{\mathcal{S}})$ preserves the distributional shape of the translation probability, including its tail, producing synthetic events that collectively cover a broader portion of the target item space. This diversity is especially useful in recommendation settings where long-tail items, which are individually low-probability but collectively important, are precisely the items that suffer most from data sparsity. We empirically validate in Section~\ref{sec:topk_vs_sampling} that sampling consistently outperforms top-$K$ selection.

\subsection{Cross-Domain Learning with Synthetic Data}
\label{sec:cross_domain_learning}

With the synthetic events $\mathcal{D}_{\text{syn}}$ generated, we now formulate the downstream training as a cross-domain learning problem. Let $f_\theta: \mathcal{U} \times \mathcal{I}_{\mathcal{T}} \rightarrow [0, 1]$ be the target domain recommendation model (e.g., a conversion prediction model) parameterized by $\theta$. The standard training objective on original target domain data is:
\begin{equation}
\mathcal{L}_{\text{orig}}(\theta) = \sum_{(u, i, y) \in \mathcal{D}_{\mathcal{T}}} \ell\!\left(f_\theta(u, i),\; y\right)
\label{eq:original_loss}
\end{equation}
where $y \in \{0, 1\}$ is the observed label and $\ell$ is a loss function (e.g., binary cross-entropy).

\begin{figure}
      \centering
      \includegraphics[width=0.6\linewidth]{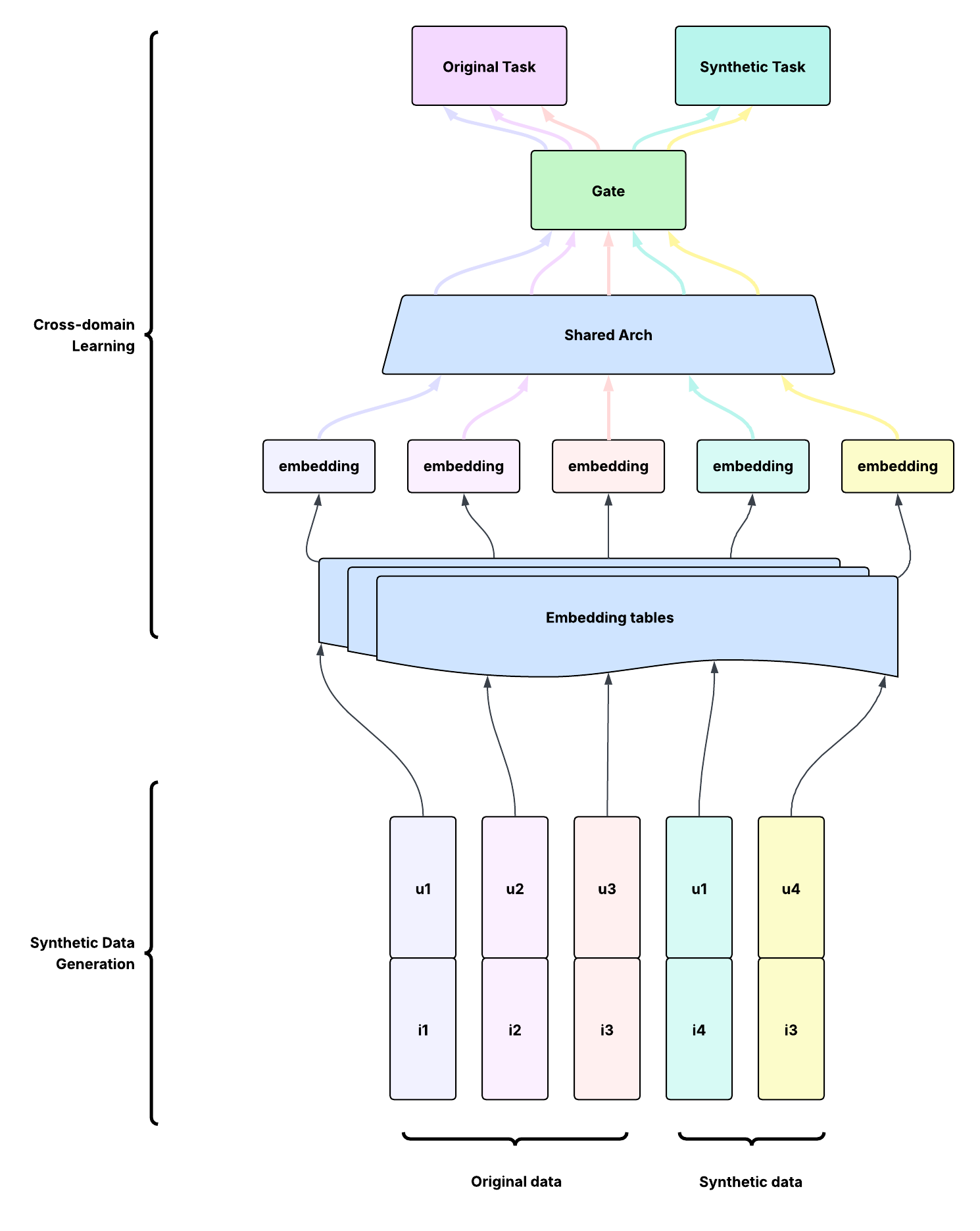}
    \caption{Illustration of model architecture for cross domain learning with Synthetic data}
\end{figure}

When training with synthetic data, we augment the objective with a weighted loss over the synthetic events:
\begin{equation}
\mathcal{L}_{\text{total}}(\theta) = \mathcal{L}_{\text{orig}}(\theta) + \lambda \sum_{(u, i, w) \in \mathcal{D}_{\text{syn}}} w \cdot \ell\!\left(f_\theta(u, i),\; \tilde{y}\right)
\label{eq:total_loss}
\end{equation}
where $\lambda > 0$ controls the relative importance of synthetic data, $w$ is the per-event confidence weight from Algorithm~\ref{alg:scalr} (uniform in our current implementation), and $\tilde{y} \in {0, 1}$ denotes the synthetic label, determined by whether a conversion event was observed in the corresponding source domain interaction.

This formulation naturally embodies cross-domain learning: the model $f_\theta$ is trained on the target domain $\mathcal{T}$, but its training signal is enriched by events \emph{originating} from the source domain $\mathcal{S}$ and \emph{translated} into the target domain's item space. The translation probabilities $\hat{P}(i_{\mathcal{T}} \mid u, j_{\mathcal{S}})$ serve as a soft alignment between domains. With our sampling-based generation and uniform weights, the cross-domain signal is encoded implicitly through the frequency of sampled items---items with higher translation probability appear more often in $\mathcal{D}_{\text{syn}}$, so their influence on training is naturally proportional to the estimated translation likelihood.

This approach differs from traditional cross-domain recommendation in that the knowledge transfer occurs entirely at the \emph{data level} rather than the \emph{model level}. The main model architecture requires no modification since the downstream model $f_\theta$ consumes synthetic events in exactly the same format as original events. The only change is the addition of an auxiliary training task (the weighted synthetic loss in Equation~\ref{eq:total_loss}), which is a minimal and architecture-independent modification. This makes SCALR applicable to any recommendation architecture (two-tower models~\cite{covington2016deep}, wide-and-deep networks~\cite{cheng2016wide}, factorization machines~\cite{guo2017deepfm}, or transformer-based models) with minimal integration effort.

\paragraph{Distribution alignment via mixing ratio.}
The hyperparameter $\lambda$ in Equation~\ref{eq:total_loss} controls the trade-off between fidelity to the original target domain distribution and the additional coverage provided by synthetic data. When $\lambda$ is too large, the model's learned distribution shifts toward the source domain's characteristics, introducing domain bias. When $\lambda$ is too small, the synthetic data contributes negligible signal. In practice, a properly tuned $\lambda$ is important for balancing the benefit of additional synthetic signal against potential domain mismatch. We leave a thorough study of the relationship between the optimal $\lambda$ and properties of the source and target domains (e.g., domain overlap, target domain sparsity, source domain volume) to future work.

% ==============================================================================
\section{Comparison with Synthetic Data in LLMs}
\label{sec:comparison}
% ==============================================================================

To better situate our work, we provide a systematic comparison between synthetic data in LLMs and in recommendation systems.

\begin{table*}[t]
\centering
\caption{Comparison of synthetic data paradigms in LLMs vs.\ recommendation systems.}
\label{tab:comparison}
\begin{tabular}{@{}lll@{}}
\toprule
\textbf{Aspect} & \textbf{LLMs} & \textbf{Recommendation Systems (Ours)} \\
\midrule
Data type & Continuous text tokens & Discrete (user, item) interaction events \\
Output space & Shared vocabulary across domains & Disjoint item catalogs across domains \\
Generation mechanism & Autoregressive language model & Conditional probability estimation \\
Quality signal & Human evaluation, perplexity & Downstream task metrics (NE, conversion rate) \\
Distribution matching & Token-level distribution & Interaction pattern distribution \\
\bottomrule
\end{tabular}
\end{table*}

Despite these differences, a fundamental principle is shared: \textbf{generating training examples that approximate the target distribution from an auxiliary source}. In LLMs, the ``auxiliary source'' is a teacher model; in our setting, it is a source domain of user interactions. The quality of the synthetic data---its fidelity to the target distribution, its diversity, and its complementarity to existing real data---determines its value in both settings.

% ==============================================================================
\section{Experiments}
\label{sec:experiments}
% ==============================================================================

We evaluate SCALR in both offline and online settings on an industrial recommendation platform.

\subsection{Experimental Setup}

\paragraph{Platform.} Our experiments are conducted on an industrial platform. The target domain $\mathcal{T}$ is conversion prediction (conversion event after seeing a recommendation). The source domain $\mathcal{S}$ consists of user engagement events from a different product surface on the same platform.

\paragraph{Scale.} The source domain contains daily interaction events. The target domain (conversion events) is orders of magnitude sparser---conversion rates are typically below 1\%. This extreme sparsity motivates the need for synthetic data augmentation.

\paragraph{Baseline.} The baseline model is a deep learning recommendation model trained on original target domain data only. We compare against the same model architecture trained on original data augmented with SCALR synthetic events.

\subsection{Online A/B Test Results}

We conducted online A/B tests over multiple weeks on live traffic. The online experiment showed consistent approximately $0.14-0.24\%$ Conversion Rate (CVR) increase across multiple weeks on one of the major models, demonstrating that the synthetic events provide useful training signal that translates to real-world performance improvements.

\subsection{Offline Analysis}
\label{sec:offline}

We investigate four research questions through controlled offline experiments:
\begin{itemize}
    \item \textbf{RQ1}: How does the generation strategy---deterministic vs.\ probabilistic---affect synthetic data quality?
    \item \textbf{RQ2}: How does synthetic data volume impact downstream performance?
    \item \textbf{RQ3}: Does aligning the synthetic data distribution to the original data improve results?
    \item \textbf{RQ4}: What other design choices influence SCALR's effectiveness?
\end{itemize}

\subsubsection{RQ1: Generation Strategy}
\label{sec:topk_vs_sampling}
A fundamental design choice in Algorithm~\ref{alg:scalr} is whether to generate synthetic events by deterministically selecting the top-$K$ items with the highest $\hat{P}(i_{\mathcal{T}} \mid u, j_{\mathcal{S}})$, or by sampling proportionally from the full distribution. We compare both strategies in the $K=1$ case: top-1 selection picks $i^* = \arg\max_i \hat{P}(i_{\mathcal{T}} \mid u, j_{\mathcal{S}})$, while probabilistic sampling draws $i^* \sim \hat{P}(\cdot \mid u, j_{\mathcal{S}})$. Both produce the same number of synthetic events, ensuring a fair volume-controlled comparison.

\begin{table}[ht]
\centering
\caption{Top-1 selection vs.\ probabilistic sampling (lower NE is better). Relative NE is reported against the original-data-only baseline.}
\label{tab:topk_vs_sampling}
\begin{tabular}{@{}lcc@{}}
\toprule
\textbf{Strategy} & \textbf{Rel.\ Train NE (\%)} & \textbf{Rel.\ Eval NE (\%)} \\
\midrule
No synthetic data & $0.00$ & $0.00$ \\
Top-1 & $-0.20$ & $-0.03$ \\
Sampling & $\mathbf{-0.26}$ & $\mathbf{-0.25}$ \\
\bottomrule
\end{tabular}
\end{table}

Top-1 selection achieves reasonable train NE ($-0.20\%$) but fails to generalize ($-0.03\%$ eval), while sampling achieves strong gains on both train ($-0.26\%$) and eval ($-0.25\%$). We attribute this to two factors. First, \textbf{diversity}: the top-1 strategy produces synthetic events covering only ${\sim}4$\% of the target item catalog, whereas sampling covers ${\sim}30$\%. Top-1 concentrates signal on a narrow set of popular items, reinforcing popularity bias rather than providing complementary coverage. Second, \textbf{distributional fidelity}: sampling preserves the shape of $\hat{P}(\cdot \mid u, j_{\mathcal{S}})$ including its tail, so the aggregate synthetic data better approximates the true cross-domain interaction pattern~\cite{holtzman2020curious,liu2024bestpractices}. Based on this finding, all subsequent experiments use probabilistic sampling.

\subsubsection{RQ2: Synthetic Data Volume}
We study the relationship between synthetic data volume and downstream performance. Starting from several thousands of synthetic rows per day, we progressively scaled to larger volumes. Performance improves monotonically as synthetic data volume increases, with diminishing returns beyond a certain scale. This suggests an optimal synthetic-to-real data ratio that balances the benefit of additional cross-domain signal against potential distribution mismatch.

\subsubsection{RQ3: Distribution Alignment}
Synthetic events tend to over-represent certain high-frequency source domain patterns. We compare three downsampling strategies for aligning the synthetic data distribution to the original training data, using relative NE on Day + 2 evaluation data:
\begin{itemize}
    \item \textbf{Bucket-wise downsampling}: Partitions synthetic events by item value buckets and downsamples each bucket to match the value distribution of the original training data.
    \item \textbf{Uniform positive downsampling}: Uniformly downsamples synthetic positive events regardless of item value, reducing overall synthetic volume to match the original data distribution.
    \item \textbf{Positive + negative downsampling}: Downsamples both synthetic positive and negative events uniformly, reducing total synthetic volume from both label classes.
\end{itemize}

\begin{table}[ht]
\centering
\caption{Impact of distribution alignment on Relative NE (lower is better), measured relative to the SCALR model without downsampling.}
\label{tab:downsample}
\begin{tabular}{@{}lc@{}}
\toprule
\textbf{Data Sampling Strategy}& \textbf{Relative NE (\%)}\\
\midrule
No downsampling (reference) & 0.00 \\
Bucket-wise downsampling & $-0.13$ \\
Uniform positive downsampling & $-0.09$ \\
Positive + negative downsampling & $-0.07$ \\
\bottomrule
\end{tabular}
\end{table}

Bucket-wise downsampling yields the largest improvement ($-0.13\%$ NE), confirming that aligning the value distribution of synthetic events to match the original data is important for downstream performance.

\subsubsection{RQ4: Additional Design Choices}
We evaluate several further design choices:
\begin{itemize}
    \item \textbf{Feature quality.} Removing features that exhibit distributional discrepancy between synthetic and real data improves eval NE, indicating that some features leak information about whether a training example is synthetic or real. Purging these features forces the model to learn from the cross-domain signal rather than exploiting artifacts.
    \item \textbf{Confidence filtering.} Filtering synthetic events based on translation confidence $\hat{P}(i_{\mathcal{T}} \mid u, j_{\mathcal{S}})$ improves eval NE, suggesting that low-confidence translations introduce noise that degrades learning.
    \item \textbf{NE--AUC divergence.} In some model variants, eval AUC improves while eval NE regresses. The model's \emph{ranking} quality benefits from synthetic signal, but predicted probabilities become over-confident, degrading calibration. This motivates the distribution alignment strategies above.
\end{itemize}

\subsection{Future Directions}

We identify several promising directions to further improve the performance of SCALR, organized into two categories.

\paragraph{Improving synthetic data quality.}
The frequency-based estimator has known limitations: it cannot handle cold-start items with no co-occurrence history, and does not model complex user-item interactions beyond pairwise co-occurrence. To address these, we have begun extending SCALR toward \emph{model-based generation}. Rather than relying on co-occurrence statistics, we train a relevance model $g_\phi$ that estimates the affinity between a user and a target item conditioned on the source item, using pre-trained embeddings:
\begin{equation}
s(u, i_{\mathcal{T}}, j_{\mathcal{S}}) = g_\phi\!\left(\mathbf{e}_u,\; \mathbf{e}_{i_{\mathcal{T}}}\right) g_\phi\!\left(\mathbf{e}_{j_{\mathcal{S}}},\; \mathbf{e}_{i_{\mathcal{T}}}\right)
\label{eq:model_based}
\end{equation}
where $\mathbf{e}_u$, $\mathbf{e}_{j_{\mathcal{S}}}$, and $\mathbf{e}_{i_{\mathcal{T}}}$ are pre-trained embeddings of the user, source item, and target item, respectively. The model is trained on observed cross-domain interactions to predict which (user, source item, target item) triples lead to engagement. At generation time, given a source event $(u, j_{\mathcal{S}})$, we sample target items from:
\begin{equation}
P_\phi(i_{\mathcal{T}} \mid u, j_{\mathcal{S}}) = \frac{\exp\!\left(s(u, i_{\mathcal{T}}, j_{\mathcal{S}}) / \tau\right)}{\sum_{i' \in \mathcal{I}_{\mathcal{T}}} \exp\!\left(s(u, i', j_{\mathcal{S}}) / \tau\right)}
\label{eq:model_sampling}
\end{equation}
where $\tau$ is a temperature parameter controlling generation diversity. This directly parallels the frequency-based $\hat{P}(i_{\mathcal{T}} \mid u, j_{\mathcal{S}})$ but replaces co-occurrence counts with learned representations, generalizing to cold-start items through embedding similarity and capturing richer user-item relationships beyond pairwise co-occurrence. In an online pretest, model-based generation achieved a 0.03\%--0.07\% increase in conversion-based revenue in a different model, validating the direction toward learned generation.

Beyond model-based generation, further improvements include \emph{addressing user behavior skewness} through user-level reweighting ($w_u = 1 / |\{(u, \cdot, \cdot) \in \mathcal{D}_{\mathcal{S}}\}|^\alpha$) or stratified sampling across activity tiers, and \emph{adversarial quality filtering} ~\cite{goodfellow2014generative,ganin2016domain} to select synthetic events most indistinguishable from real target domain events.

\paragraph{Extending the generation framework.}
The current framework generates synthetic events based on observed source domain interactions, that is, events that actually happened. Generating synthetic \emph{counterfactual} events which represent interactions that a user would likely \emph{not} engage in could provide additional contrastive signal during model training. Additionally, incorporating \emph{temporal alignment} between source and target domain events (e.g., restricting generation to recent source interactions versus leveraging full history) may further improve the relevance of synthetic data.

% ==============================================================================
\section{Discussion}
\label{sec:discussion}
% ==============================================================================

\paragraph{Why ``synthetic data'' rather than ``cross-domain recommendation''?}
While SCALR inherently addresses cross-domain recommendation, its conceptual framing is fundamentally distinct.  Classical cross-domain recommendation has primarily focused on \emph{model-level} transfer on sharing parameters, embeddings, or representations across domains~\cite{zhu2021crossdomain}. In contrast, SCALR introduces a \emph{data-level} transfer paradigm by generating new training events that augment the target domain's dataset. This data-centric perspective aligns with the broader trend in machine learning toward data-centric AI~\cite{liu2024bestpractices} and enables a modular, model-agnostic approach. Furthermore, explicitly connecting this approach to the synthetic data paradigm in LLMs allows us to adapt extensive best practices developed in that domain, such as quality filtering, distribution matching, and the management of synthetic-to-real data ratios into the recommendation setting.

\paragraph{Broader implications.}
If synthetic data proves as transformative for recommendation systems as it has for LLMs, it could fundamentally change how recommendation models are trained. Rather than being limited to directly observed interactions, recommendation systems could systematically leverage the full spectrum of user behavior across platforms and surfaces, translating diverse signals into domain-specific training data. This vision connects to the broader goal of building unified user models that integrate information across all touchpoints.

% ==============================================================================
\section{Conclusion}
\label{sec:conclusion}
% ==============================================================================

We have presented SCALR, a framework for synthetic data learning in recommendation systems that generates training events for a target domain by translating user interactions from a source domain. By explicitly framing cross-domain event transfer as synthetic data generation and drawing parallels to the transformative role of synthetic data in LLM training, we provide a principled, modular, and model-agnostic approach to data augmentation for recommendation systems.

Our frequency-based estimation method serves as an efficient and scalable baseline, and our online experiments on a large-scale industrial platform demonstrate statistically significant improvements in conversion prediction. The unique challenges of synthetic data in recommendation systems including user behavior skewness, discrete heterogeneous item spaces, and cross-domain intent shift differentiate this problem from synthetic data generation in LLM and open rich directions for future research.

We believe that synthetic data learning represents an important and underexplored paradigm for recommendation systems, with the potential to address fundamental challenges of data sparsity, long-tail distribution, and cold-start---much as synthetic data has addressed data scarcity in LLM training.

%%
%% The acknowledgments section is defined using the "acks" environment
%% (and NOT an unnumbered section). This ensures the proper
%% identification of the section in the article metadata, and the
%% consistent spelling of the heading.
% \begin{acks}
% To Robert, for the bagels and explaining CMYK and color spaces.
% \end{acks}

%%
%% The next two lines define the bibliography style to be used, and
%% the bibliography file.
\bibliographystyle{assets/plainnat}
\bibliography{references}

%%
%% If your work has an appendix, this is the place to put it.
% \appendix

\end{document}